 \documentclass[final,3p,11pt]{elsarticle}

\usepackage{natbib}
\biboptions{numbers,sort&compress}
\usepackage[nodots]{numcompress}
\usepackage{graphicx}
\usepackage{caption}
\usepackage{subcaption}
\usepackage{amsfonts,amssymb,amsbsy,amsmath,mathtools}

\journal{Composite Structures}

\begin{document}

\begin{frontmatter}

\title{A nonlinear couple stress-based sandwich beam theory}
\tnotetext[label0]{This is only an example}

 \author[add1]{Bruno Reinaldo Goncalves\corref{cor1}}
\ead{bruno.reinaldogoncalves@aalto.fi}
\cortext[cor1]{Corresponding author. Tel.: +358453308058}
\address[add1]{Aalto University, Department of Mechanical Engineering, FI-00076 Aalto, Finland}
\address[add2]{Texas A\&M University, Department of Mechanical Engineering, College Station, TX 77843-3123, USA}
\author[add1,add2]{Anssi T. Karttunen}
\author[add1]{Jani Romanoff}

\begin{abstract}
A geometrically nonlinear sandwich beam model founded on the modified couple stress Timoshenko beam theory with \textit{von K\'arm\'an} kinematics is derived and employed in the analysis of periodic sandwich structures. The constitutive model is based on the mechanical behavior of sandwich beams, with the bending response split into membrane-induced and local bending modes. A micromechanical approach based on the structural analysis of a unit cell is derived and utilized to obtain the stiffness properties of selected prismatic cores. The model is shown to be equivalent to the classical thick-face sandwich theory for the same basic assumptions. A two-node finite element interpolated with linear and cubic shape functions is proposed and its stiffness and geometric stiffness matrices are derived. Three examples illustrate the model capabilities in predicting deflections, stresses and critical buckling loads of elastic sandwich beams including elastic size effects. Good agreement is obtained throughout in comparisons with more involved finite element models.
\end{abstract}


\end{frontmatter}


\section{Introduction}
\label{sec1}

Sandwich panels are lightweight structures composed of two faces separated by a low-density core. Over the past decades, these structures have found numerous applications where high stiffness-to-weight/strength ratios are important, such as in vehicle engineering. The two faces are relatively thin and stiff, whereas the core is comparatively thick and soft. The face and core materials as well as the core topology can be tailored based on the desired mechanical properties of the assembly \cite{steeves2004,fleck2010}. In recent years, technological advances such as aluminium brazing, laser-welding and additive manufacturing have allowed the production of sandwich structures with highly optimized periodic cores, see Fig.~1 and Refs. \cite{lokcheng2000,romanoff2007,stpierre2015} for examples. Continuum modelling is an efficient way to predict their response without discretely modelling all structural details involved. \par

Continuum models for sandwich structures can be broadly divided into single-layer and layerwise categories according to the dependency of layer-level variables \cite{carrera2008}. Oftentimes, refinements of classical theories are referred to as higher-order theories, such as the works in \cite{he2012,frostig1990}. Low-complexity single-layer models are used in applications where their through-thickness behavior assumptions hold approximately. In particular, first-order shear deformation theories have been used extensively to analyze sandwich beams and plates coupled with standard mechanics or homogenization approaches \cite{lokcheng2000,buannic2003}. Conventional first-order theories are, however, inaccurate near discontinuities such as point forces or restrictive boundary conditions. They assume that cell and structure scales are clearly separated and hence that local effects are negligible. Still, periodic sandwich structures can display scale interactions in the presence of relatively large, flexible cells. To include such interactions, enhanced single-layer sandwich theories have been proposed under different assumptions. In particular, the thick-face sandwich theory \cite{allen1969,plantema1966,zenkert1995,romanoff2007} and other enhanced models that include the effect of thick faces \cite{romanoff2007a,jelovica2018} have been used to analyze periodic sandwich structures. In recent years, non-classical continuum theories have gained footing in the study of sandwich structures. For instance, couple stress \cite{reddy2011,ma2008,arbind2013,yang2002} and micropolar theories are suitable for the analysis of periodic sandwich beams \cite{romanoff2014,goncalves2017,goncalves2018,karttunen2018a} and other lattice structures \cite{liu2009,noor1980,bazant1972} involving different levels of complexity and accuracy. \par

In this work, we develop a couple stress-based model for the analysis of sandwich beams. The model is founded on the modified couple stress Timoshenko beam theory \cite{ma2008,reddy2011} and utilizes a constitutive matrix tailored based on the mechanical behavior of sandwich structures. A micromechanical approach based on the structural analysis of a unit cell is defined and utilized to determine the stiffness properties of selected periodic sandwich beams. The model is shown to match the thick-face sandwich beam theory \cite{allen1969,plantema1966,zenkert1995} in the linear case for the same basic assumptions. A two-node finite element is formulated and interpolated using linear Lagrangian and cubic Hermitian shape functions, and the underlying stiffness and geometric stiffness matrices are derived. Three examples validate the theory against more involved finite element models and compare the results with the conventional Timoshenko beam theory and thick-face sandwich theory. The examples concern linear and geometrically nonlinear bending and linear buckling of periodic sandwich beams with different core shear flexibility levels.

\begin{figure}[h]
\centering
\includegraphics[scale=1]{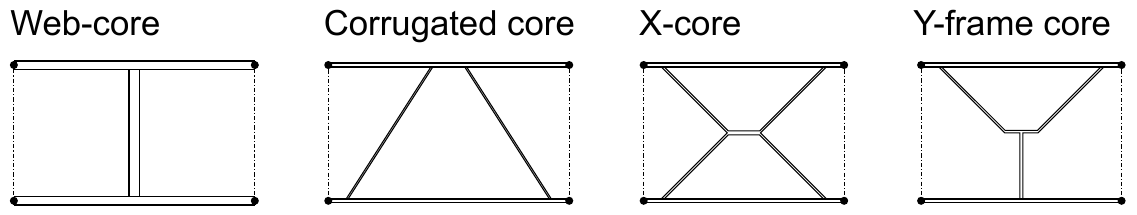}
\caption{Unit cells of periodic sandwich beams with applications in civil, mechanical and vehicle engineering.}
\end{figure} 

\begin{figure}[h]
\centering
\includegraphics[scale=1]{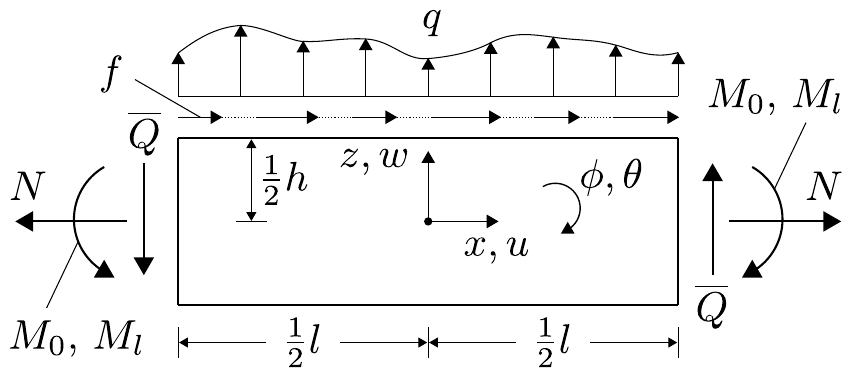}
\caption{Couple stress sandwich beam model with conventions indicated.}
\end{figure} 

\section{Couple stress-based sandwich theory}
\label{sec2}

\subsection{Couple stress beam model}

Let us consider the modified couple stress theory \cite{yang2002, ma2008, reddy2011} to construct an equivalent sandwich Timoshenko beam model that includes the local bending effects arising near discontinuities such as point loads. Figure 2 shows the general conventions of a couple stress  beam of length $l$ and height $h$. The displacement field of the beam can be written as
\begin{equation}
U_{x}(x,z) = u(x)+z\phi(x), \quad U_{z}(x,z) = w(x)
\end{equation}
where $w$ and $\phi$ are the transverse displacement and the cross-sectional rotation about the $y$-axis, respectively. The nonzero strains including the \textit{von K\'arm\'an} nonlinearity are \cite{reddy2011}
\begin{equation}
\begin{aligned}
\epsilon_x&=\frac{\partial u}{\partial x}+\frac{1}{2}\Big(\frac{\partial w}{\partial x}\Big)^2+z\frac{\partial \phi}{\partial x}=\epsilon_0+z\kappa_\phi, \\ 
\gamma_{xz}&=\phi-\theta, \\  
\chi_{xy}&=\frac{1}{4}\Big(\frac{\partial \phi}{\partial x}-\frac{\partial^2 w}{\partial x^2}\Big) = \frac{1}{4}(\kappa_\phi+\kappa)
\end{aligned}
\end{equation}
where $\theta = -\frac{\partial w}{\partial x}$, and $\chi_{xy}$ is the curvature related to the macrorotation. The beam model at hand can transmit couple stress $m_{xy}$, as well as the usual normal stress $\sigma_x$ and the transverse shear stress $\tau_{xz}$. The axial $N$, periodic shear $Q_0$ and two independent bending stress resultants $M_0$ and $M_l$ are defined (Fig~2)
\begin{equation}
\begin{aligned}
N = \int_{A}^{} \sigma_xdz, \quad Q_0 = K_s\int_{A}^{} \sigma_{xz}dz, \quad M_0 = \int_{A}^{}  \Big(\sigma_xz+\frac{1}{2}m_{xy}\Big)dz, \quad M_l = \frac{1}{2}\int_{A}^{} m_{xy}dz
\end{aligned}
\end{equation}
Using the conventions here proposed and following the general steps in \cite{reddy2011}, the equilibrium equations are obtained for the static case in absence of applied body couples
\begin{equation}
\begin{aligned}
\frac{\partial N}{\partial x}+f=0, \quad
\frac{\partial \overline{Q}}{\partial x}+q=0, \quad
Q_0-\frac{\partial M_0}{\partial x}=0, \quad
\end{aligned}
\end{equation}
where $f$ and $q$ are, respectively, distributed axial and transverse loads, and
\begin{equation}
\begin{aligned}
\overline{Q} = Q_0 + Q_l + N\frac{\partial w}{\partial x}, \quad Q_l = \frac{\partial M_l}{\partial x}
\end{aligned}
\end{equation}
The boundary conditions, defined at $x=\pm \ l/2$, are
\begin{equation}
\begin{aligned}
N \ \ \textrm{or} \ \ u, \quad \overline{Q} \ \ \textrm{or} \ \ w, \quad M_0 \ \ \textrm{or} \ \ \phi, \quad M_l \ \ \textrm{or} \ \ \theta 
\end{aligned}
\end{equation}

\begin{figure}[h]
\centering
\includegraphics[scale=1]{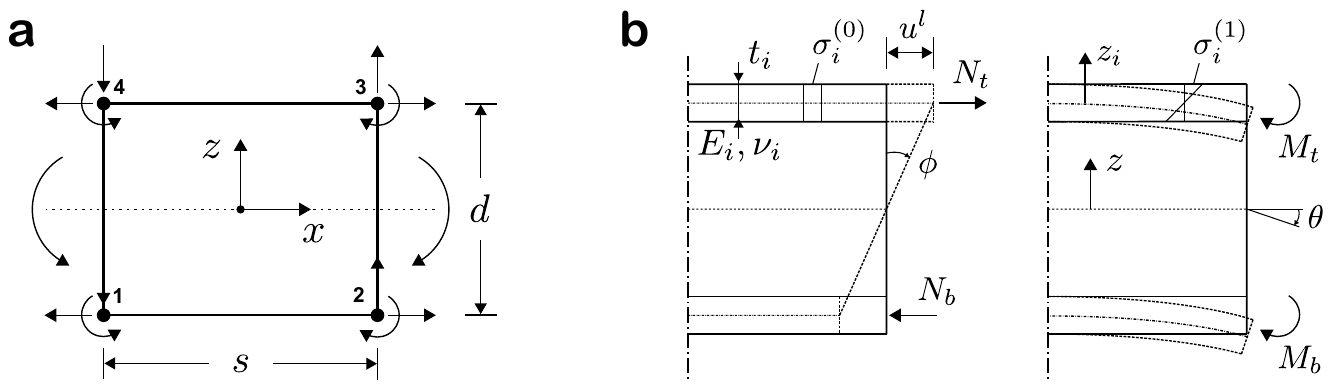}
\caption{(a) General arrangement of a sandwich beam unit cell (b) Membrane stretch and local bending of sandwich members.}
\end{figure} 

\subsection{Sandwich constitutive model}
Consider in Fig.~3a a unit-width sandwich cell composed of two continuous faces and arbitrary periodic core, with all members assumed to behave as conventional Euler-Bernoulli beam elements. The vertical cell limits are defined at the centerline of the faces, whose distance characterizes the depth $d$. The unit cell boundaries at $x = \pm \ s/2$ are taken as void except at its four corners, see examples in Fig.~1. The couple stress beam resultants translate into equivalent forces and moments at the corner nodes \textbf{1} to \textbf{4}.  A constitutive model is defined with coupling between normal stress and local curvature, and couple stress and axial strain
\begin{equation}
\begin{aligned}
\sigma_x = Q_{11}\epsilon_x + Q_{13}\chi_{xy}, \quad \sigma_{xz}=Q_{22}\gamma_{xz}, \quad 2m_{xy}= Q_{31}\epsilon_x+Q_{33}\chi_{xy}
\end{aligned}
\end{equation}
where $Q_{13}=Q_{31}$ results from coordinate system invariance of the constitutive equations and symmetry of the stress tensor. Two independent bending modes are, in general, identified. The first, related to the area term of the parallel axis theorem, is the sandwich effect, while the second is the local bending of the prismatic members to the total curvature (Fig~3b). According to the elementary beam theory, the corresponding member-level stresses can be written 
\begin{equation}
\begin{aligned}
\sigma_i^{(0)} = E_i \frac{\partial u^l}{\partial x} = E_i(z-z_i(z))\frac{\partial \phi}{\partial x}, \quad \sigma_i^{(1)} = - E_iz_i(z)\frac{\partial^2 w^l}{\partial x^2}, \quad z_i = z-z_i^0, \ w^l = w
\end{aligned}
\end{equation}
where $z_i$ is the vertical coordinate of the local coordinate system with origin at $z_i^0$. We then define the bending-inducing stresses of the couple stress beam based on the discrete member-level response
\begin{equation}
\begin{aligned}
\sigma_i^{(0)}z \equiv \sigma_xz + \frac{1}{2}m_{xy}, \quad \sigma_i^{(1)}z \equiv \frac{1}{2}m_{xy}
\end{aligned}
\end{equation}
Let us expand force- and bending-inducing stresses of the couple stress beam
\begin{equation}
\begin{aligned}
\sigma_x &=  Q_{11} \epsilon_0 + \Big(Q_{11}z+\frac{Q_{13}}{4}\Big)\kappa_\phi + \frac{Q_{13}}{4}\kappa \\
\sigma_xz + \frac{1}{2}m_{xy} &= \Big(Q_{11}z+\frac{Q_{13}}{4}\Big)\epsilon_0 + \Big(Q_{11}z^2+\frac{Q_{13}}{2}z + \frac{Q_{33}}{16}\Big)\kappa_\phi + \Big(Q_{13}z + \frac{Q_{33}}{4}\Big)\kappa \\
\frac{1}{2}m_{xy} &= \frac{Q_{13}}{4}\epsilon_0 + \Big(\frac{Q_{13}z}{4}+\frac{Q_{33}}{16}\Big)\kappa_\phi + \frac{Q_{33}}{16}\kappa
\end{aligned}
\end{equation}
Substituting Eq.~(9) into Eq.~(10) and knowing that the modes are uncoupled, the following relationships are obtained
\begin{equation}
\begin{aligned}
\frac{Q_{33}}{16} = E_iz_iz, \quad Q_{11}z^2+\frac{Q_{13}}{2}z + \frac{Q_{33}}{16} = E_i(z^2-zz_i), \quad \frac{Q_{13}z}{4}+\frac{Q_{33}}{16} = 0, 
\end{aligned}
\end{equation}
Solving the system of equations, the constitutive terms reduce to 
\begin{equation}
\begin{aligned}
Q_{11} = E(z), \quad Q_{22} = G(z), \quad Q_{13} = - 4E(z)z_i(z), \quad Q_{33} = 16E(z)zz_i(z)
\end{aligned}
\end{equation}
where $Q_{22}$ is obtained following the same reasoning as in the conventional Timoshenko beam theory. We substitute the $Q_{ij}$ terms into Eq.~(10) and further into Eq.~(3), acknowledging that $\int E(z)z_idz = 0$, to obtain the relations between stress resultants and displacement gradients, which define the beam constitutive matrix $\textbf{C}$
\begin{equation}
\begin{Bmatrix}
N \\
M_0 \\
Q_0 \\
M_l
\end{Bmatrix}
=
\begin{bmatrix}
A & B  & 0      & 0 \\
B & D_0  & 0      & 0\\
0      &    0    & D_Q & 0 \\
0      &    0    & 0      & D_l \\
\end{bmatrix}
\begin{Bmatrix}
\epsilon_0 \\
\kappa_\phi \\
\gamma_{xz} \\
\kappa 
\end{Bmatrix}
\end{equation}
where $D_0 = D_g-D_l$. Note that the local bending moment does not induce axial strain as result of the thickness-averaging process, i.e. $C_{14} = C_{41} = 0$. The axial $A$, bending $D_g, D_l$, and axial-bending coupling $B$ terms of a sandwich beam are defined
\begin{equation}
\begin{aligned}
A &= \int_{z}^{} E(z)dz, \quad B = \int_{z}^{} E(z)zdz, \quad D_g = \int_{z}^{} E(z)z^2dz, \\
D_l &= k_c\int_{z}^{} E(z)zz_i(z)dz, \quad D_Q = k_s\int_{z}^{} G(z)dz
\end{aligned}
\end{equation}
where $A, B, D_g, D_Q$ are equal to the terms in the conventional Timoshenko beam theory, while $D_l$ is the additional term due to the local cell bending stiffness. The coefficient $k_l$ describes the non-uniform distribution of couple stresses over the depth. In the present work, we assume that couple stresses are approximately uniform, thus $k_c = 1.0$.

\section{Micromechanical approach for periodic sandwich structures}
\label{sec3}

Let us define an alternative, straightforward approach to determine stiffness for highly discrete sandwich beams, for which direct through-thickness integration as in Eq.~(14) can become complex. Consider a periodic cell following the conventions of Fig.~2 and Fig.~3a and modelled using conventional, nodally-exact Euler-Bernoulli beam elements. Boundary displacements are applied to the corners inducing deformation modes $\Delta$, which are utilized to determine the constitutive terms in Eq.~(13) for a sandwich beam with arbitrary periodic core. In addition to the displacement boundary conditions discussed next, the cell should be constrained to prevent rigid-body motion.

\begin{figure}[h]
\centering
\includegraphics[scale=1]{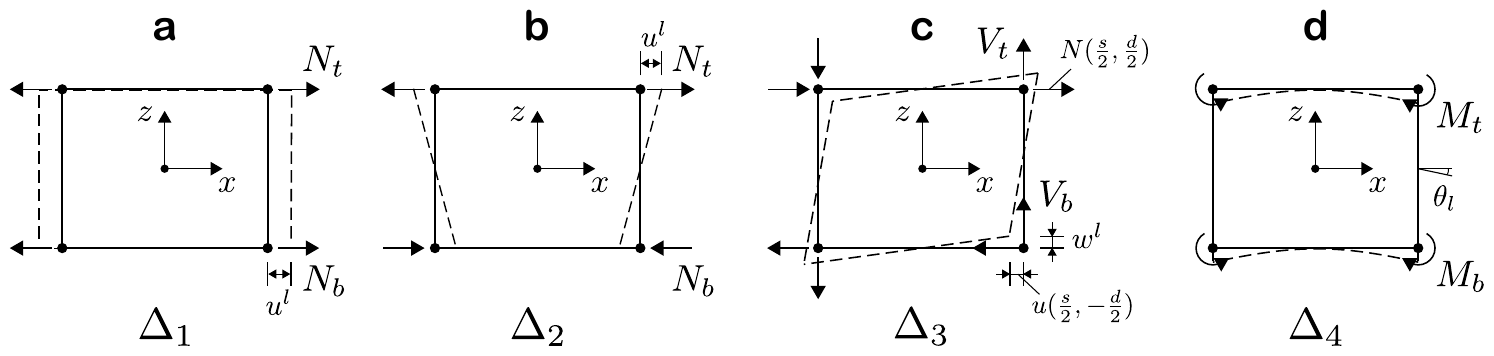}
\caption{Displacement boundary conditions used in the micromechanical approach for stiffness computation.}
\end{figure}

\subsection{Axial, bending and coupling terms}
Figures~4a-b show idealized strain cases that involve axial deformation of the sandwich faces and core. The displacement boundary conditions that define $\Delta_1$ and $\Delta_2$ are
\begin{equation}
\begin{aligned}
&\Delta_1: \quad  \ u^l = u\Big(\frac{s}{2},z\Big)=-u\Big(-\frac{s}{2},z\Big), \\
&\Delta_2: \quad  \ u^l = u\Big(\frac{s}{2},\frac{d}{2}\Big) = -u\Big(\frac{s}{2},-\frac{d}{2}\Big),  \quad u\Big(\frac{s}{2},z\Big)=-u\Big(-\frac{s}{2},z\Big)
\end{aligned}
\end{equation}
The equivalent non-zero strain components are
\begin{equation}
\begin{aligned}
\Delta_1: \quad \epsilon_0=\frac{2u^l}{s}, \quad \Delta_2: \quad \kappa_\phi=\frac{2u^l}{sd}
\end{aligned}
\end{equation}
Based on the conventions of Fig.~(2) and Eq.~(3), the stress resultants are obtained
\begin{equation}
N=N_t+N_b, \quad M_0=(N_t-N_b)d
\end{equation}
The combinations that define the axial, bending and axial-bending stiffness terms become
\begin{equation}
\begin{aligned}
\Delta_1: \quad A&=\frac{N}{\epsilon_0}=\frac{(N_t+N_b)s}{2u^l}, \quad B=\frac{M_0}{\epsilon_0}=\frac{(N_t-N_b)sd}{2u^l}, \\ 
\Delta_2: \quad B&=\frac{N}{\kappa_\phi}=\frac{(N_t+N_b)sd}{2u^l}, \quad D_0=\frac{M_0}{\kappa_\phi}=\frac{(N_t-N_b)sd^2}{2u^l}
\end{aligned}
\end{equation}
where $B$ obtained with either model can be shown numerically to be equal.

\subsection{Transverse shear stiffness}
Consider in Fig.~4c a sandwich cell under periodic transverse shear deformation. The boundary conditions that define this mode are given by
\begin{equation}
\begin{aligned}
&w^l = w\Big(\frac{s}{2},z\Big)=-w\Big(-\frac{s}{2},z\Big), \quad   u\Big(\frac{s}{2},z\Big)=u\Big(-\frac{s}{2},z\Big), \\
&N\Big(\frac{s}{2},\frac{d}{2}\Big) = -N\Big(-\frac{s}{2},\frac{d}{2}\Big)  = -N\Big(\frac{s}{2},-\frac{d}{2}\Big) = N\Big(-\frac{s}{2},-\frac{d}{2}\Big) = \frac{2(V_t+V_b)}{sd}
\end{aligned}
\end{equation}
The equivalent transverse shear strain and resultant $Q_0$ are given by
\begin{equation}
\begin{aligned}
\gamma_{xz}=\frac{2w^l}{s}+\frac{1}{d}\Big[u\Big(\frac{s}{2},\frac{d}{2}\Big)-u\Big(\frac{s}{2},-\frac{d}{2}\Big)\Big], \quad Q_0=V_t+V_b,
\end{aligned}
\end{equation}
whose quotient defines $D_Q$
\begin{equation}
\begin{aligned}
D_Q=\frac{Q_0}{\gamma_{xz}}=\frac{(V_t+V_b)sd}{2w^ld+\Big[u(\frac{s}{2},\frac{d}{2})-u(\frac{s}{2},-\frac{d}{2})\Big]s}
\end{aligned}
\end{equation}

\subsubsection{Local bending term}
Figure 4d shows an unit cell composed of flexural-only elements under constant curvature. The displacement boundary conditions, strain and stress resultants that define $\Delta_4$ are given by 
\begin{equation}
\begin{aligned}
\theta^l = \theta\Big(\frac{s}{2},z\Big) = -\theta\Big(-\frac{s}{2},z\Big), \quad \kappa=\frac{2\theta^l}{s}, \quad M_l=M_t+M_b
\end{aligned}
\end{equation}
The local bending term $D_l$ is defined
\begin{equation}
\begin{aligned}
D_l&=\frac{(M_t+M_b)s}{2\theta^l}
\end{aligned}
\end{equation}
which is equivalent to the length-average bending stiffness of the cell prismatic members.

\section{Selected periodic sandwich cores}
\label{sec4}

Figure~5 shows the unit cells of simple periodic sandwich beams discretized with Euler-Bernoulli beam elements. The faces (1)-(4) have equal thickness $t_f$, while the core members (5),(6) have thickness $t_c$. The material properties are equal for all elements, described by the elastic constants $E$ and $\nu$. The stiffness terms of Eq.~(13) are determined by enforcing the displacement conditions of the micromechanical model to nodes \textbf{1}-\textbf{4}, evaluating the boundary resultants and computing the relations of Eq.~(18), (21) and (23). Node \textbf{5} has all degrees of freedom constrained to prevent rigid-body motion. In addition to the stiffness properties, discrete resultants at the sandwich faces are determined based on the homogeneous beam solution.

\begin{figure}[h]
\centering
\includegraphics[scale=1]{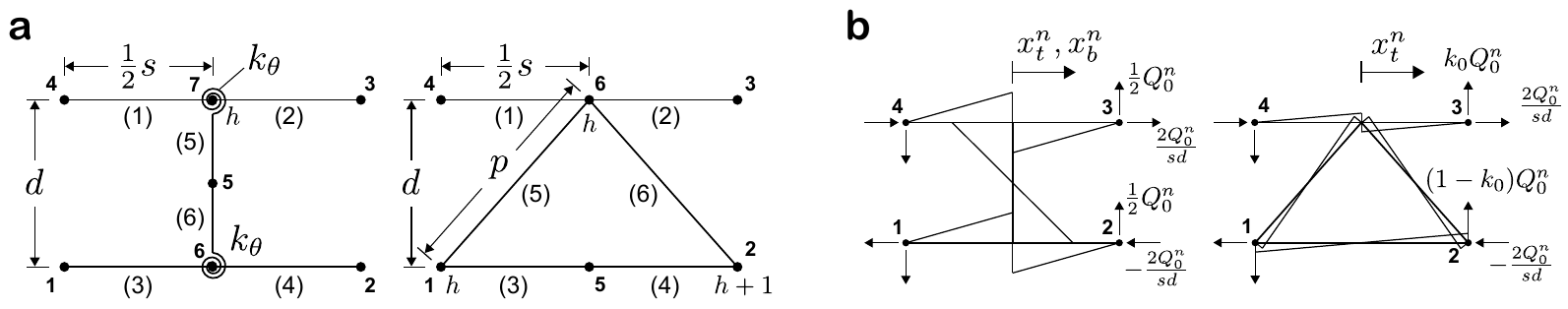}
\caption{(a) Conventions used in the analysis of web-core and triangular corrugated core cells (b) Periodic shear-induced moment distribution.}
\end{figure} 

\subsection{Web-core with semi-rigid joints}
The axial, bending and axial-bending stiffness terms for the web-core cell in Fig.~5a are
\begin{equation}
A=2Et_f, \quad B=0, \quad D_0= \frac{1}{2}Et_fd^2, \quad D_l = \frac{Et_f^3}{6}
\end{equation}
and the transverse shear stiffness, which results in the deformation mode in Fig.~5b for a web-core beam with semi-rigid face-core joints of stiffness $k_\theta$, is given by
\begin{equation} 
D_Q=\frac{2Et_f^3t_c^3}{s(k_1t_f^3t_c^3+2dt_f^3+st_c^3)}, \quad k_1 = \frac{E}{k_\theta}
\end{equation}

\subsection{Triangular corrugated core}
The axial, bending and axial-bending stiffness terms for a triangular corrugated cell as in Fig.~5a are given by
\begin{equation}
A=E\Big(2t_f+\frac{st_c^3}{8pd^2}\Big), \quad B=-\frac{Est_c^3}{16pd}, \quad D_0= E\Big(\frac{t_fd^2}{2}+\frac{st_c^3}{32p}\Big), \quad D_l = E\Big(\frac{t_f^3}{6}+\frac{st_c^3}{24p}\Big)
\end{equation}
and the transverse shear stiffness
\begin{equation}
D_Q = \frac{4Esd^2t_ft_c}{8p^3t_f+s^3t_c}
\end{equation}

\subsection{Discrete response at the sandwich faces}
Let us define a simplified discrete stress analysis scheme for the selected structures based on the sandwich continuum quantities. We assume that $t_c^2 \ll d^2$ for the triangular corrugated core, and exclude the nonlinear term in Eq.~(2)a. The interval between two consecutive hard points is defined $n = [h,h+1]$ (Fig.~5a). The membrane force and bending moment at top and bottom faces, $i = t,b$, result for both sandwich beams
\begin{equation} 
N_i^n \approx \frac{1}{s}\int_{x^h_i}^{x^{h+1}_i} \Big( \frac{N}{2} \pm \frac{M_0}{d} \Big) \ dx, \quad M_i^n \approx \frac{D_{l,i}}{D_l}\frac{1}{s}\int_{x^h_i}^{x^{h+1}_i} M_l \ dx + M_{Q}^n(x)
\end{equation}
where $x_i^h$ is the $x$-coordinate of the hard point $h$ at the face $i=t,b$. The shear-induced bending moment is given by
\begin{equation} 
M_{Q}^n = (k_0Q_0^n+k_lQ_l^n) \Big( x_i^n - \frac{s}{2} \Big), \quad Q_0^n = \frac{1}{s}\int_{x^h_i}^{x^{h+1}_i} Q_0 \ dx, \quad Q_l^n = \frac{1}{s}\int_{x^h_i}^{x^{h+1}_i} Q_l \ dx
\end{equation}
where $x_i^n=x-x_i^h$ and $k_0, k_l$ define, respectively, the share of periodic and local shear forces to which each face is subjected. The factor $k_0$ is obtainable from the boundary conditions in Eq.~(19). For the web-core cell of Fig.~5, $k_0 = k_l = 1/2$, while $k_l \approx k_0$ is assumed for the corrugated cell. The discrete stress for the sandwich faces $i = t,b$ within the interval $n$ is given by
\begin{equation} 
\sigma_i^n(x_i^n,z_i) = \frac{N_i^n}{t_f} + \frac{12M_i^nz_i}{t_f^3}
\end{equation}
where $z_i$ is the local vertical coordinate consistent with the definition of Fig.~3b.

\section{Couple stress and thick-face sandwich theory equivalence}
\label{sec5}

\subsection{Thick-face sandwich beam theory}

We revise the thick-face sandwich beam theory according to the conceptual framework of Allen \cite{allen1969} and Plantema \cite{plantema1966}. In their works, the effect of thick faces is included by studying the shear deformation compatibility between core and faces. The global sandwich beam response is defined
\begin{equation}
q_1 =-\frac{\partial Q_1}{\partial x}, \quad Q_1=\frac{\partial M_1}{\partial x}, \quad M_1=-D_g\frac{\partial^2 w_1}{\partial x^2}
\end{equation}
Near discontinuities, the faces must bend to a finite curvature for faces and core to remain attached. Thus, they are locally subjected to a set of loads, shear forces and bending moments
\begin{equation}
q_2 =-\frac{\partial Q_2}{\partial x}, \quad Q_2=\frac{\partial M_2}{\partial x}, \quad M_2=-D_f\frac{\partial^2 w_2}{\partial x^2}
\end{equation}
The corresponding total quantities are given by
\begin{equation}
q = q_1+q_2, \quad Q = Q_1+Q_2, \quad M = M_1+M_2, \quad w = w_1+w_2
\end{equation}
Shear strain compatibility between face and core results in the following relations
\begin{equation}
-Q_1 = D_g\frac{\partial^3 w_1}{\partial x^3} = -D_Q\frac{\partial w_2}{\partial x}+D_f\frac{\partial^3 w_1}{\partial x^3}
\end{equation}
which after some rearranging becomes
\begin{equation}
\frac{\partial^2 Q_1}{\partial x^2}-a^2Q_1=-a^2Q, \quad a^2 = \frac{D_QD_g}{D_fD_0}
\end{equation}
Eq.~(35) can be adapted for linear buckling analysis of sandwich beams. Consider the presence of a bending-inducing axial force $P$; the total shear force becomes $Q = P (\frac{\partial w_1}{\partial x}+\frac{\partial w_2}{\partial x})$. Acknowledging that $\frac{\partial w_2}{\partial x}=\frac{Q_1}{a^2D_f}$, the buckling equilibrium equation becomes
\begin{equation}
\frac{\partial^5 w_1}{\partial x^5}-\Big(a^2-\frac{P}{D_f}\Big)\frac{\partial^3 w_1}{\partial x^3}-\frac{a^2P}{D_g}\frac{\partial w_1}{\partial x}=0
\end{equation}

\subsection{Equivalence between models}

Let us introduce some general assumptions to study the equivalence between the thick-face sandwich theory and the couple stress sandwich model
\begin{itemize}
   \item The axial degree of freedom $u$ of the couple stress beam is removed;
  \item The sandwich beam has an antiplane core. That is, the normal stress at the core is zero (i.e. $E_c=0)$ and the cross-sectional shear stress distribution is constant over the core thickness;
  \item Horizontal sliding, denoted  $\gamma_0$ in Allen \cite{allen1969}, is excluded from the analyses;
\end{itemize}
as a result of the first assumption, the equilibrium equations of the couple stress beam result
\begin{equation}
\begin{aligned}
\frac{\partial Q_0}{\partial x}+\frac{\partial Q_l}{\partial x}-q=0, \quad Q_0-\frac{\partial M_0}{\partial x}=0
\end{aligned}
\end{equation}
Based on the second assumption and considering the faces to be equal, we obtain the following common stiffness terms for both, thick-face and couple stress sandwich models
\begin{equation}
\begin{aligned}
D_0 &= \frac{E_ft_fd^2}{2}, \quad D_l = D_f = 2E_fI_f, \quad D_g = D_0+D_l
\end{aligned}
\end{equation}
We shall now proceed with the equivalency derivations, which are developed for the static bending case. Substituting the relations in Eq.~(13) into Eq.~(37b), we obtain a relation between shear angle and cross-sectional rotation angle
\begin{equation}
\begin{aligned}
\gamma_{xz}=\frac{D_0}{D_Q}\frac{\partial^2 \phi}{\partial x^2}
\end{aligned}
\end{equation}
Writing Eq.~(37a) in terms of displacements, and substituting the shear angle definition of Eq.~(2b)
\begin{equation}
\begin{aligned}
q = D_g\frac{\partial^3 \phi}{\partial x^3}-D_f\frac{\partial^3 \gamma_{xz}}{\partial x^3}
\end{aligned}
\end{equation}
Integrating Eq.~(37a) and defining $\int qdx = Q$, we obtain
\begin{equation}
\begin{aligned}
Q = D_g\frac{\partial^2 \phi}{\partial x^2}-D_f\frac{\partial^2 \gamma_{xz}}{\partial x^2} + C = Q_g+Q_l^\gamma
\end{aligned}
\end{equation}
where the constant of integration obtained is included in $Q_g$. We now substitute $Q_g$ into Eq.~(39) and differentiate twice
\begin{equation}
\begin{aligned}
\frac{\partial^2 \gamma_{xz}}{\partial x^2}=\frac{D_0}{D_QD_g}\frac{\partial^2 Q_g}{\partial x^2}
\end{aligned}
\end{equation}
Isolating $\frac{\partial^2 \gamma_{xz}}{\partial x^2}$ in Eq.~(41) and substituting in Eq.~(42), the following differential equation is obtained
\begin{equation}
\begin{aligned}
\frac{D_fD_0}{D_gD_Q}\frac{\partial^2 Q_g}{\partial x^2}= -Q+Q_g \rightarrow{} \frac{\partial^2 Q_g}{\partial x^2}-a^2Q_g=-a^2Q, \quad a^2 = \frac{D_QD_g}{D_fD_0}
\end{aligned}
\end{equation}
Eq.~(43) is equal to the governing equation of the thick-face sandwich theory, Eq.~(35), acknowledging that $Q_g = Q_1$. Therefore, the two theories are shown to be equivalent for the basic assumptions of \cite{allen1969}.

\section{Couple stress finite element model}
\label{sec6}

Consider the finite element of length $l_e$ and height $h_e$ shown in Fig.~6. The element has two nodes and four degrees of freedom per node
\begin{equation}
\begin{aligned}
\mathbf{u}=\left\{u_1 \ \ \phi_1 \ \ w_1 \ \ \theta_1 \ \ u_2 \ \ \phi_2 \ \ w_2 \ \ \theta_2 \right\}^{\textrm{T}} \end{aligned}
\end{equation}
with positive directions following the conventions in Fig.~2. We approximate the primary variables $u$ and $\phi$ using Lagrange linear polynomials $\psi_i$, while $w$ and $\theta$ are approximated using Hermitian cubic polynomials $\varphi_i$
\begin{equation}
\begin{aligned}
u(x) = \sum_{i=1}^{2}u_i\psi_i, \quad \phi(x) = \sum_{i=1}^{2}\phi_i\psi_i, \quad w(x) = \sum_{i=1}^{4}\overline{\Delta}_i\varphi_i
\end{aligned}
\end{equation}
where
\begin{equation}
\begin{aligned}
u_1 = u(-l_e/2), \quad u_2 = u(l_e/2), \quad \phi_1 = \phi(-l_e/2), \quad \phi_2 = \phi(l_e/2)\\
\overline{\Delta}_1 = w(-l_e/2), \quad \overline{\Delta}_2 = \theta(-l_e/2), \quad \overline{\Delta}_3 = w(l_e/2), \quad \overline{\Delta}_4 = \theta(l_e/2),
\end{aligned}
\end{equation}
The stress resultants at the nodes  are determined in relation to the positive directions of Fig.~1
\begin{equation}
\begin{aligned} 
N_1 &= - N(-l_e/2), \quad \overline{Q}_1 = -\overline{Q}(-l_e/2), \quad M_{0,1} = -M_0(-l_e/2), \quad M_{l,1} = -M_l(-l_e/2) \\
N_2 &= N(l_e/2),  \quad \overline{Q}_2 = \overline{Q}(l_e/2), \quad M_{0,2} = M_0(l_e/2), \quad M_{l,2} = M_l(l_e/2) \\
\end{aligned}
\end{equation}
We present next the finite element equations for linear, geometric nonlinear and eigenvalue buckling analyses.

\begin{figure}[h]
\centering
\includegraphics[scale=1]{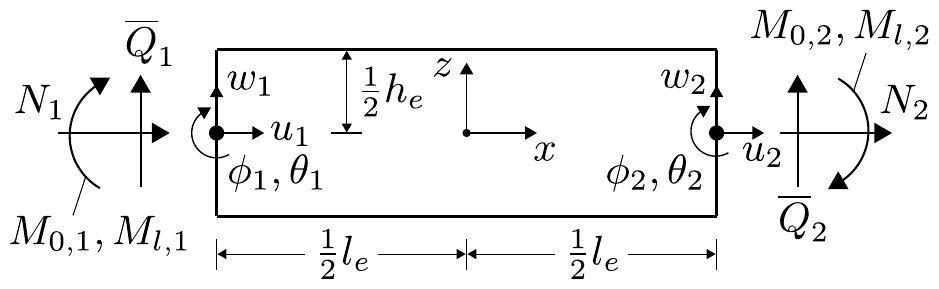}
\caption{Two-node beam element for the couple stress sandwich beam theory.}
\end{figure}

\subsection{Linear finite element equations}
The linear finite element equations arise from removing the \textit{von K\'arm\'an} term from the axial strain description. The displacements are linearly proportional to the applied load, and the principle of superposition is valid. In this case, the elemental stiffness matrix becomes
\begin{equation}
\begin{aligned}
\mathbf{K}=\int_{-\frac{l_e}{2}}^{\frac{l_e}{2}} \mathbf{B}^T\mathbf{C}\mathbf{B} \ dx \
\end{aligned}
\end{equation}
where $\mathbf{B}$ (Eq.~(A.1)) is the linear strain-displacement matrix and $\mathbf{C}$ is the constitutive matrix in Eq.~(13). The standard finite element equations $\mathbf{F} = \mathbf{K}\mathbf{u}$ are then solved after enforcing the loads and boundary conditions. 

\subsection{Nonlinear finite element equations}
Let us now consider the \textit{von K\'arm\'an} nonlinear problem. In an analogy with Eq.~(48), the updated finite element equations become \cite{kant1992,wood1977}
\begin{equation}
\mathbf{K}=\int_{-\frac{l_e}{2}}^{\frac{l_e}{2}}(\mathbf{B}+\theta \mathbf{B}_\sigma)^T\mathbf{C}(\mathbf{B}+\frac{\theta}{2}\mathbf{B}_\sigma) \ dx 
\end{equation}
where $\mathbf{B}_\sigma$ is defined in Eq.~(A.2). In short, the objective is to minimize the residual
\begin{equation}
\mathbf{R}=\mathbf{K(u)}\mathbf{u}-\mathbf{F}
\end{equation}
using a solution procedure for nonlinear differential equations such as the Newton-Raphson method. The tangent stiffness matrix derivations follow the standard steps as in \cite{reddy2015}.

\subsection{Linear buckling analysis}
Let us now consider the eigenvalue elastic buckling problem. Assume that the beam is subjected to a constant axial force $P$ that induces transverse displacements. Following the basic steps in \cite{cook1989}, the geometric stiffness matrix becomes 
\begin{equation}
\begin{aligned}
\mathbf{K}_\sigma=\int_{-\frac{l_e}{2}}^{\frac{l_e}{2}} P \ \mathbf{B_\sigma}^T\mathbf{B_\sigma} \ dx 
\end{aligned}
\end{equation}
The conventional eigenvalue buckling problem is then solved
\begin{equation}
\begin{aligned}
(\mathbf{K}-\lambda\mathbf{K}_\sigma)\mathbf{d}_\sigma = 0
\end{aligned}
\end{equation}
where the eigenvalues $\lambda$ correspond to the buckling loads and the eigenvector $\mathbf{d}_\sigma$ provides the buckling modes.

\section{Numerical results}
\label{sec7}

\subsection{General assumptions}

In order to demonstrate the couple stress-based sandwich beam theory (CSS), we study the linear and geometric nonlinear response of periodic sandwich beams. In the following analyses, the beams represent wide panels along the \textit{y}-axis, whose response is two-dimensional. Plane strain conditions are then assumed, with elastic modulus set to $E=E_s/(1-\nu_s^2)$. The following material properties are taken: $E_s = 206$ GPa and $\nu_s = 0.3$. The results are shown for an unit-width beam. The examples are validated using finite element models (3D FE) that represent the 3-D geometry (Fig.~7), constructed with Abaqus S4R shell elements. In the validation models, vertical point forces are applied to the sandwich structural hard points at the relevant $x$-coordinate. Comparisons with the thick-face sandwich theory (TFS) and the conventional Timoshenko beam (TBT) with effective properties are shown. 

\begin{figure}[h]
\centering
\includegraphics[scale=1]{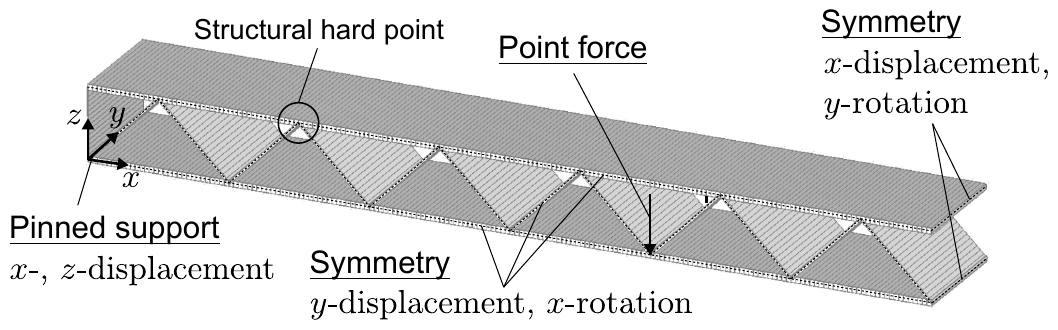}
\caption{Example of three-dimensional FE model with common load and boundary condition assumptions as used for validation.}
\end{figure} 

\subsection{Stress analysis of periodic sandwich beams}

Let us consider web-core and triangular-corrugated core sandwich beams of equal core density ($\approx 4.7 \%$) and cell dimensions shown in Fig.~8. The web-core has semi-rigid face-core joints with the average rotational stiffness reported in \cite{romanoff2007c}. Twelve cell repetitions along the \textit{x}-axis define the structures, to a total of $L=0.72m$. The beams are subjected to four-point bending within a single linear, quasi-static step, to a total vertical displacement of $-0.001m$ at $x=0.24m$ ($L/3$) and symmetry conditions at mid-length. Figure 8 shows the vertical displacement distributions along the beam axes using the couple stress sandwich theory, thick-face sandwich theory \cite{allen1969,plantema1966,zenkert1995} and validation models (bottom face). Good agreement is observed between couple stress and validation models, whereas couple stress and thick-face beams predict equal displacements.

\begin{figure}[h]
\centering
\includegraphics[scale=1]{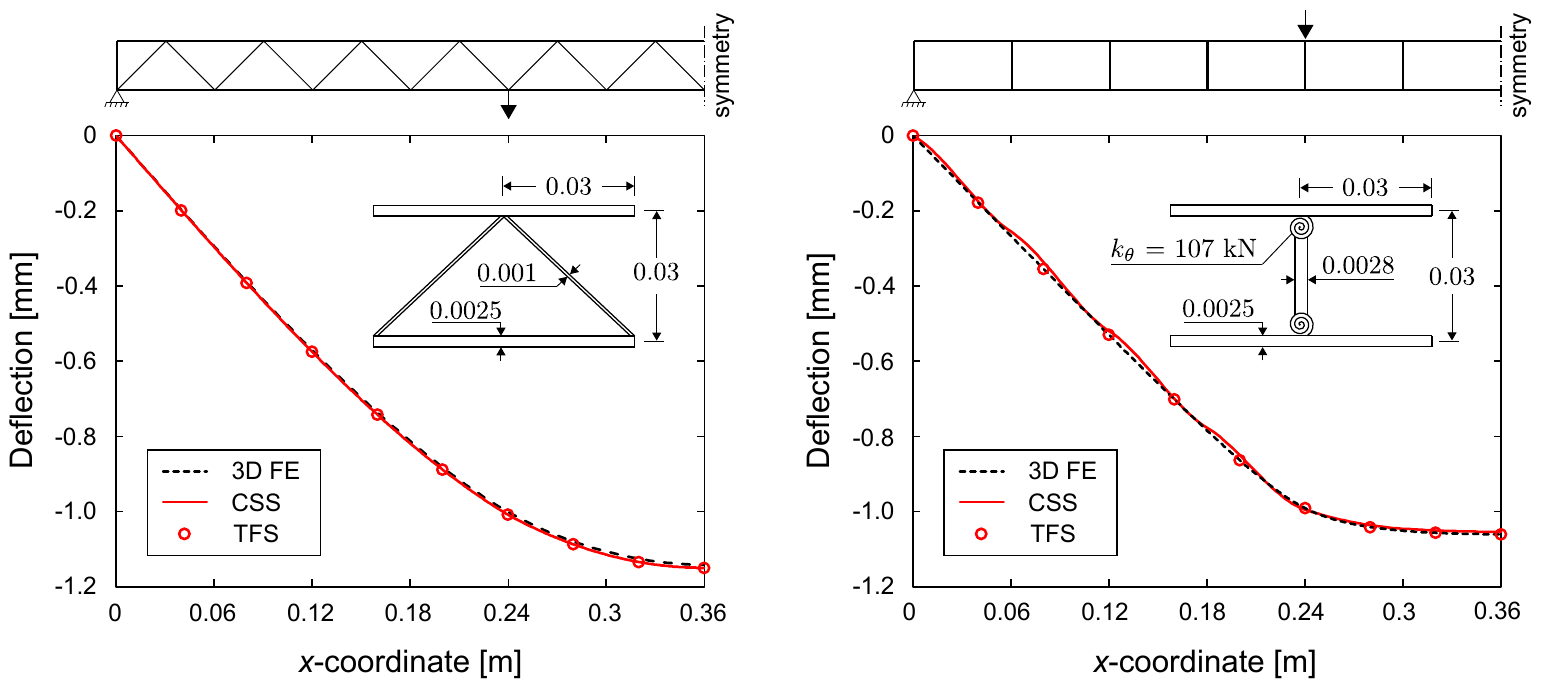}
\caption{Deflections obtained with couple stress sandwich (CSS), thick-face sandwich (TFS) and 3D finite element (3D FE) models for (a) triangular corrugated and (b) web-core sandwich beams in four-point bending. Cell dimensions are shown in [m].}
\end{figure}

\begin{figure}[]
\centering
\includegraphics[scale=1]{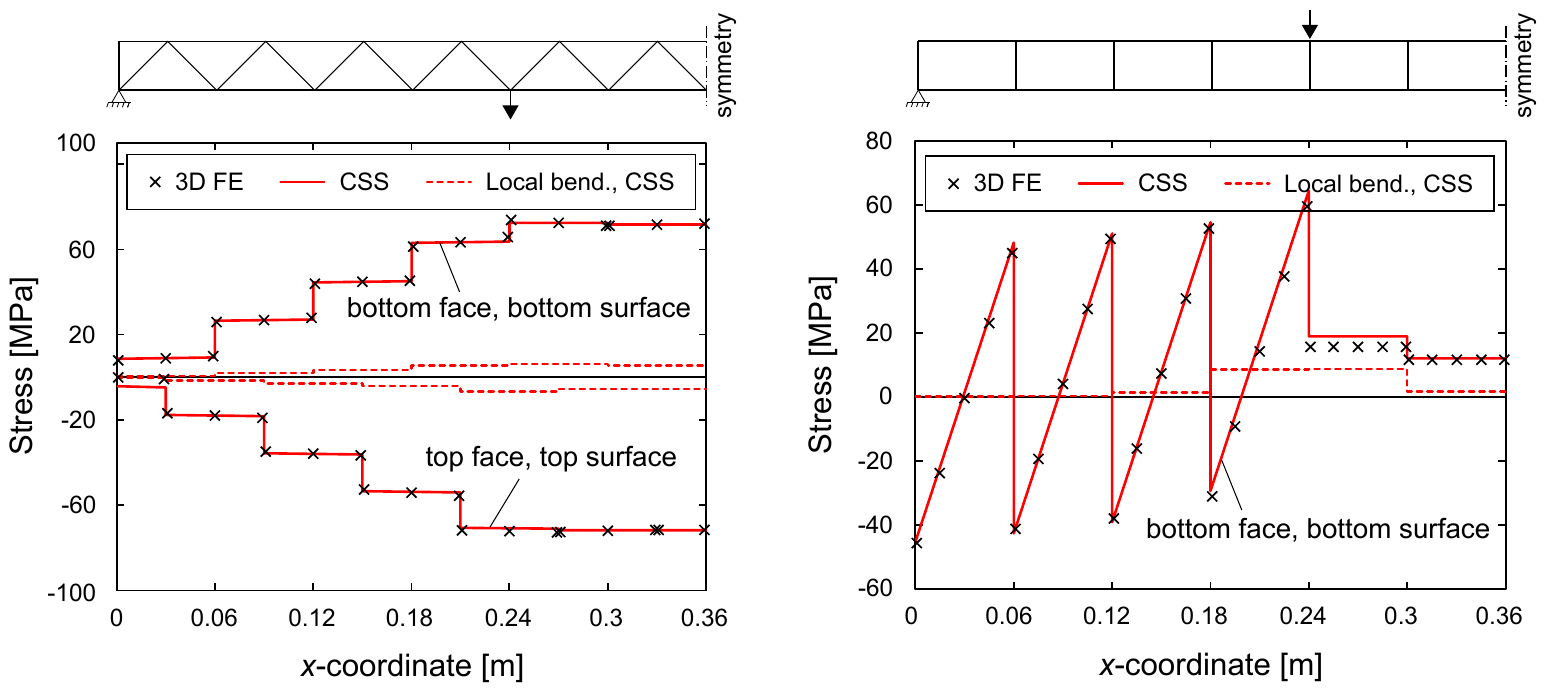}
\caption{Stress distributions of (a) triangular corrugated (top and bottom faces) and (b) web-core sandwich (bottom face) beams in four point bending as obtained from the couple stress sandwich beam model.}
\end{figure} 

Figure 9a shows the bottom surface stress at the bottom face and top surface stress at the top face of the triangular corrugated core beam. Figure 9b shows the bottom surface stress at the bottom face of the web-core beam; the top face stress is qualitatively similar due to mid-depth symmetry. Local bending stresses are also presented in Fig.~9. Stresses are localized from the homogeneous solution as described in detail in Section 4.3. Overall, the couple stress sandwich beam is able to predict the stress distributions with good accuracy against the validation model. Local bending stresses are comparatively higher in the web-core structure at hand due to its higher shear flexibility. Yet, it is shown that local bending has a non-negligible contribution for an accurate stress analysis of both structures.

\begin{figure}[]
\centering
\includegraphics[scale=1]{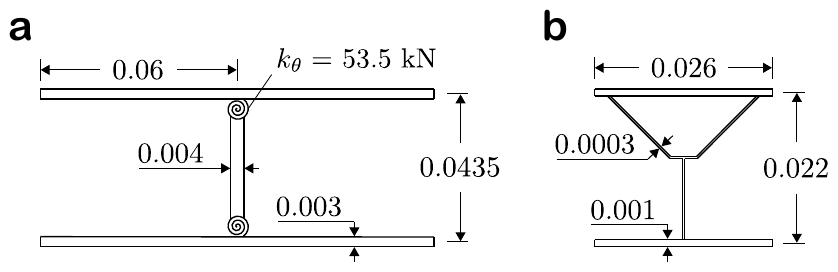}
\caption{Unit cells of web-core and Y-frame core sandwich beams used in Sections 7.3 and 7.4. Dimensions in [m].}
\end{figure}

\subsection{Nonlinear bending of web-core sandwich beams}

Consider now in Fig.~10a a web-core unit cell with semi-rigid joints, whose rotational stiffness is half of the average reported in \cite{romanoff2007c}. A sandwich beam composed of 10 of such cells ($L=1.2m$) is subjected to bending. Two doubly-clamped settings are investigated, where the load is either concentrated at the mid-length, or distributed over the top face plate. The Newton-Raphson algorithm is utilized in conjunction with the finite element method to obtain an approximate solution to the nonlinear equilibrium equations. The convergence tolerance is set to $10^{-4}$ and the analysis is divided into fine load steps. Figure 8 shows the load vs. maximum deflection relations in either case, as well as comparisons with the conventional Timoshenko beam and validation model. Overall, the couple stress sandwich theory satisfactorily predicts the nonlinear response in terms of displacements. In the linear range, a slight error (4-5\%) related to the periodic shear description of the couple stress model is observed. The example under consideration is an extreme case with relatively few, highly shear-flexible cells. More involved models, such as the micropolar theory \cite{karttunen2018a}, can be used in such case for a more precise response prediction. The error is rapidly reduced as the geometric nonlinearity increases and the membrane action becomes dominant. In both cases, the couple stress sandwich model is considerably more accurate than the conventional Timoshenko beam theory with effective properties. 

\begin{figure}[]
\centering
\includegraphics[scale=.95]{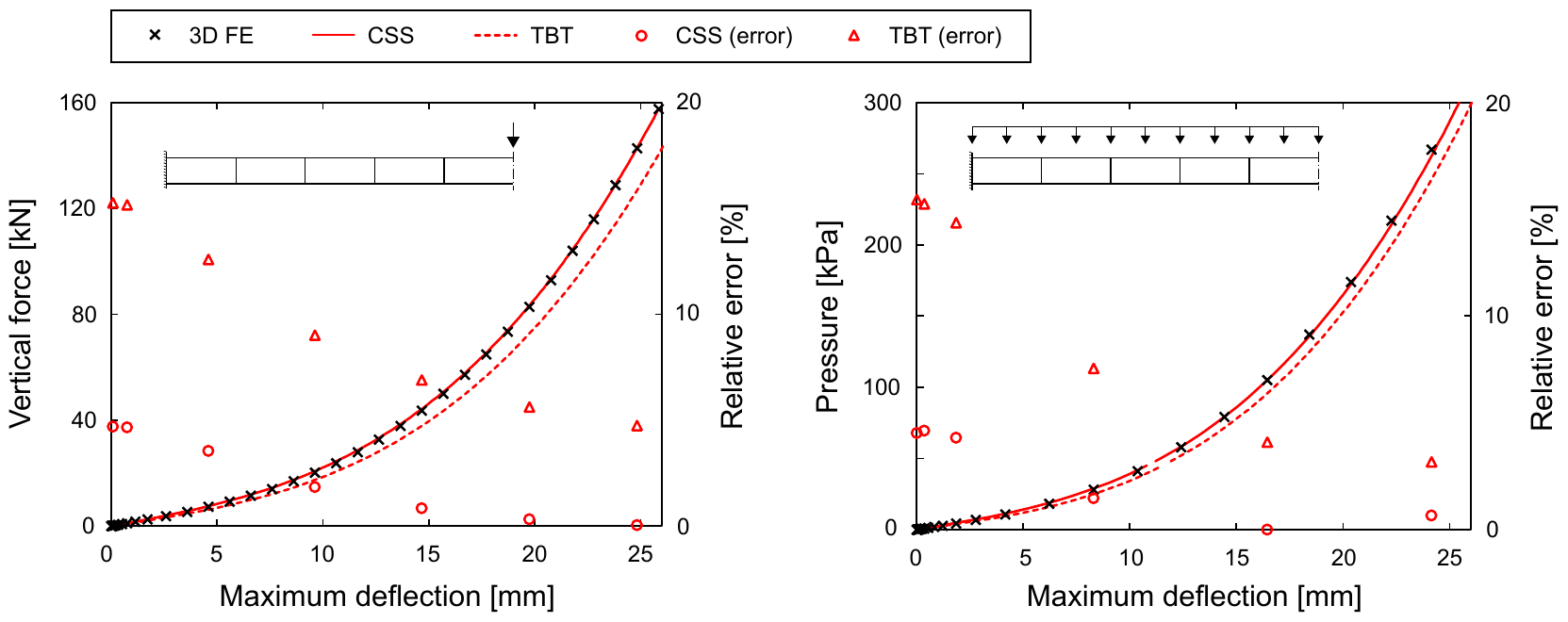}
\caption{Nonlinear bending analysis of clamped web-core sandwich beams under (a) centered point force (b) distributed load.}
\end{figure} 

\begin{figure}[]
\centering
\includegraphics[scale=.95]{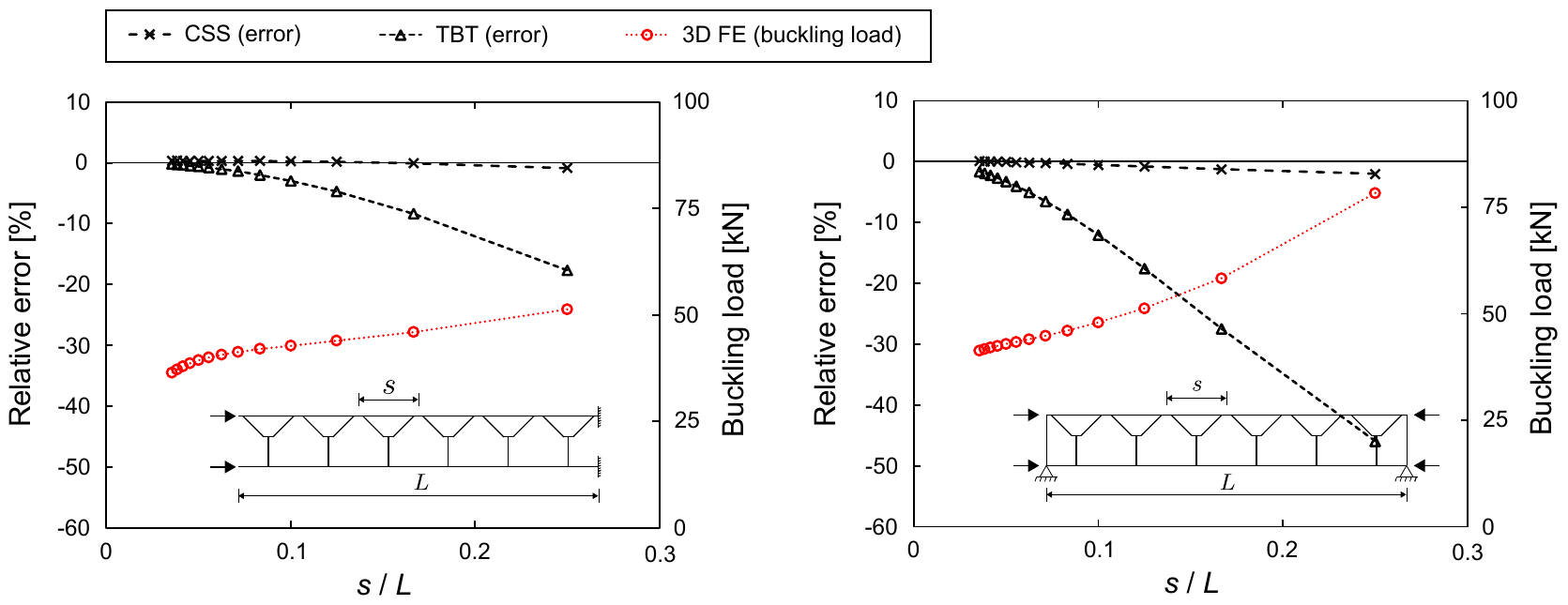}
\caption{Elastic buckling load of Y-frame sandwich structures and relative error in the predictions using the couple stress sandwich and classical Timoshenko models.}
\end{figure} 

\subsection{Elastic buckling of Y-frame core sandwich beams}

Figure~10b shows the simplified unit cell of a Y-frame core sandwich beam with similar dimensions as in \cite{stpierre2015}. Y-frame structures used in ship design are often composed of relatively few cells, thus prone to size effects, which we investigate next. We analyze the changes in linear buckling loads of axially compressed Y-frame beams as function of their relative unit-cell size $s/L$. Two configurations are covered, namely an end-loaded cantilever and a pinned-pinned beam loaded at both ends. Figure 11 shows the buckling load predicted with the validation models and the relative errors obtained with classical Timoshenko and couple stress sandwich models. Overall, the conventional Timoshenko model is progressively less accurate as the cell size approaches the total structural length. This supports the observations in \cite{goncalves2017,goncalves2018}, where a conventional couple stress beam was used to model the sandwich structures. The couple stress sandwich beam succeeds in describing size effects in Y-frame core sandwich structures, displaying good accuracy against the more involved validation models.

\section{Conclusions}
A sandwich beam model founded on the modified couple stress Timoshenko beam theory has been defined and employed in the analysis of elastic periodic sandwich structures. A standard micromechanical approach has been proposed as a generalization of previous works (e.g. \cite{libove1951,goncalves2018, lokcheng2000, sun1996}) for the couple stress sandwich model, and applied to determine effective stiffness properties of selected cores. The model was shown to be equivalent to the thick-face sandwich theory in the linear case \cite{allen1969,plantema1966,zenkert1995} for the same basic assumptions. Unlike the thick-face model, however, the couple stress sandwich beam relies on a single kinematical definition to describe the scales involved. The kinematical variables are equal to the conventional Timoshenko beam theory except of a higher-order curvature term. \par

The model herein derived improves the single-layer description of sandwich beams when compared to the conventional Timoshenko model. It is able to predict the structural behaviour near discontinuities such as point loads or clamped boundary conditions, and thus capture size effects. When compared to the modified couple stress Timoshenko model for layered structures \cite{ma2008,reddy2011}, the redefinition of resultants and introduction of coupling constitutive coefficients guarantee that the model describes the deformation of a sandwich beam as in Refs. \cite{allen1969,zenkert1995}. The stress resultants have a direct correspondence to the sandwich beam member-level forces and moments, which facilitates stress localization for any periodic core. Stress predictions were shown to be reasonably accurate for selected bending- and stretch-dominated cores, which are respectively more and less prone to size effects \cite{goncalves2018}. \par

\section*{Acknowledgements}
The authors gratefully acknowledge the financial support from the Graduate School of Aalto University School of Engineering.

\appendix

\section{Finite element matrices}
\label{appendix-sec1}
The elemental matrices needed for the finite element computations are given as follows. The linear strain-displacement matrix $\mathbf{B}$ that approximates the relations between nodal displacements (Eq.~(44)) and linear strains of the couple stress sandwich model is given by
\begin{equation}
\mathbf{B}=
\begin{bmatrix}
\psi_1^{'} & 0 & 0 & 0 & \psi_2^{'} & 0 & 0 & 0 \\
0           & \psi_1^{'} & 0 & 0 & 0 &\psi_2^{'} & 0 & 0\\
0           &  \psi_1    & \varphi_1^{'} & -\varphi_2^{'} & 0 &\psi_2 & \varphi_3^{'} & -\varphi_4^{'} \\
0           &   0        & -\varphi_1^{''} & \varphi_2^{''} & 0 & 0 & -\varphi_3^{''} & \varphi_4^{''}\\
\end{bmatrix} 
\end{equation}
The geometric strain-displacement matrix, which interpolates the \textit{von K\'arm\'an} nonlinear terms, is given by 
\begin{equation}
\mathbf{B}_\sigma=
\begin{bmatrix}
0 & 0 &-\varphi_1^{'} & \varphi_2^{'} & 0 & 0 &-\varphi_3^{'} & \varphi_4^{'} \\
0 & 0 & 0 & 0 & 0 & 0 & 0 & 0 \\
0 & 0 & 0 & 0 & 0 & 0 & 0 & 0 \\
0 & 0 & 0 & 0 & 0 & 0 & 0 & 0 \\
\end{bmatrix} 
\end{equation}



\bibliographystyle{elsarticle-num}
\bibliography{sample}

\end{document}